\documentclass{sf2a-conf2011}
\usepackage{graphicx}
\usepackage{hyperref}
\usepackage[]{natbib}
\usepackage[cyr]{aeguill}
\usepackage{epstopdf}

\def\BibTeX{{\rm B\kern-.05em{\sc i\kern-.025em b}\kern-.08em
    T\kern-.1667em\lower.7ex\hbox{E}\kern-.125emX}}
\bibpunct{(}{)}{;}{a}{}{,}  

\begin{document}

\TitreGlobal{SF2A 2011}


\title{The Galactic Center region viewed by H.E.S.S.}

\runningtitle{The Galactic Center region viewed by H.E.S.S.}

\author{A. Viana$^{1,}$}\address{CEA, Irfu, Centre de Saclay, F-91191 Gif-sur-Yvette | France}\address{on behalf of the H.E.S.S. collaboration}




\setcounter{page}{237}

\index{Viana, A.}


\maketitle


\begin{abstract}
The Galactic center region is the most active region in the Milky Way harboring a wealth of photon sources at all wavelengths. H.E.S.S. observations of the Galactic Center (GC) region revealed for the first time in very high energy (VHE, E> 100 GeV) gamma-rays a detailed view of the innermost 100 pc of the Milky Way and provided a valuable probe for the acceleration processes and propagation of energetic particles near the GC. H.E.S.S. has taken more than 180 hours of good-quality observations toward the GC region since the experience started in 2003. A strong and steady gamma-ray source has been detected coincident in position with the supermassive black hole Sgr A*. Besides the central pointlike source, a diffuse emission extended along the Galactic Plane has been detected within about 1$^{\circ}$ around the GC. An accurate analysis of the Galactic center region suggests that the diffuse emission may dominate highest energy end of the overall GC source spectrum. I will review the current VHE view by H.E.S.S. of the GC region and briefly discuss the theoretical models which explain VHE gamma-ray emissions of the central source and the diffuse emission.
\end{abstract}



\section{Introduction}
  The Galactic Center (GC) region harbours a variety of potential sources of high-energy radiation
including the supermassive black hole Sagittarius (Sgr) A* of $2.6 \times 10^6$ M$_{\odot}$~\citep{Schodel:2002py},  and a number of supernova remnants, among them the Sgr A East remnant of a giant supernova explosion which happened about 10000 years ago. The Galactic Center was therefore a prime target for observations with Imaging Atmospheric Cherenkov telescopes (IACTs), and detection of very high energy (VHE, E> 100 GeV) gamma rays was reported by The CANGAROO~\citep{Tsuchiya:2004wv}, VERITAS~\citep{Kosack:2004ri}, H.E.S.S.~\citep{Aharonian:2004wa} and MAGIC\citep{Albert:2005kh} from the direction of the Galactic Center (GC). The nature of this source is still unknown. The H.E.S.S. observations of the GC region led to the detection of a point-like source of VHE gamma-rays at the gravitational center of the Galaxy (HESS J1745-290), compatible with the positions of the supermassive black hole Sgr A*, the supernova remnant (SNR) Sgr A East, and the plerion G359.95-0.04. A larger exposure of the region in 2004 revealed a second source: the supernova remnant G0.9+0.1~\citep{Aharonian:2005br}. The subtraction of these two sources revealed a ridge of diffuse emission extending along the Galactic plane for roughly 2$^{\circ}$ (Fig.~\ref{viana:fig2}).

\section{The H.E.S.S. instrument}
The H.E.S.S. (High Energy Stereoscopic System) experiment is an array of four identical imaging atmospheric Cherenkov telescopes located in the Khomas Highland of Namibia ($23^{\circ} 16^{\prime} 18^{\prime \prime}$ South, $16^{\circ}30^{\prime}00^{\prime \prime}$ East) at an altitude of 1800~m above sea level. Each telescope has an optical reflector consisting of 382 round facets of 60~cm diameter each, yielding a total mirror area of 107~m$^2$. The Cherenkov light, emitted by charged particles in the electromagnetic showers initiated by primary gamma rays, is focused on cameras equipped with 960 photomultiplier tubes, each one subtending a field-of-view of $0.16^\circ$. The large field-of-view ($\sim$$5^\circ$) permits survey coverage in a single pointing. The direction and the energy of the primary gamma rays are reconstructed by the stereoscopic technique.

\section{HESS J1745-290: counterparts and spectrum}

 In December 2004, H.E.S.S. reported the detection of VHE gamma rays from the center of our Galaxy, at the time based on data obtained with the first two H.E.S.S. telescopes during 16h of observations in 2003. Within the - at the time unprecedented - precision of 30" in RA and Dec, the location of the source HESS J1745-290 was consistent with the Galactic gravitational center, and the spectrum of gamma rays was consistent with a power law up to 10 TeV. Towards identifying the origin of the gamma rays, a multi-year effort was invested aimed at improving the pointing position of the H.E.S.S. telescopes. After a careful investigation of the pointing systematics of the H.E.S.S. telescopes, the systematic error on the centroid position was reduced from 30" to 6" per axis, with a comparable statistical error - by far the best source location achieved in gamma rays so far~\citep{2010MNRAS.402.1877A}. The thus determined source position is within 8"$\pm$9"$_{\rm stat} \pm$9"$_{\rm sys}$ from Sgr A*, well consistent with the location of the black hole and the pulsar wind nebula (PWL) G359.95-0.04, but it excludes Sgr A East remnant  as the main counterpart of the VHE emission at the level of 5-7$\sigma$, depending on the assumed position of the VHE emission in Sgr A East~\citep[Fig. \ref{viana:fig1}, left; see][for more details]{2010MNRAS.402.1877A}.
\begin{figure}[t!]
 \centering
 \includegraphics[width=0.65\textwidth,clip]{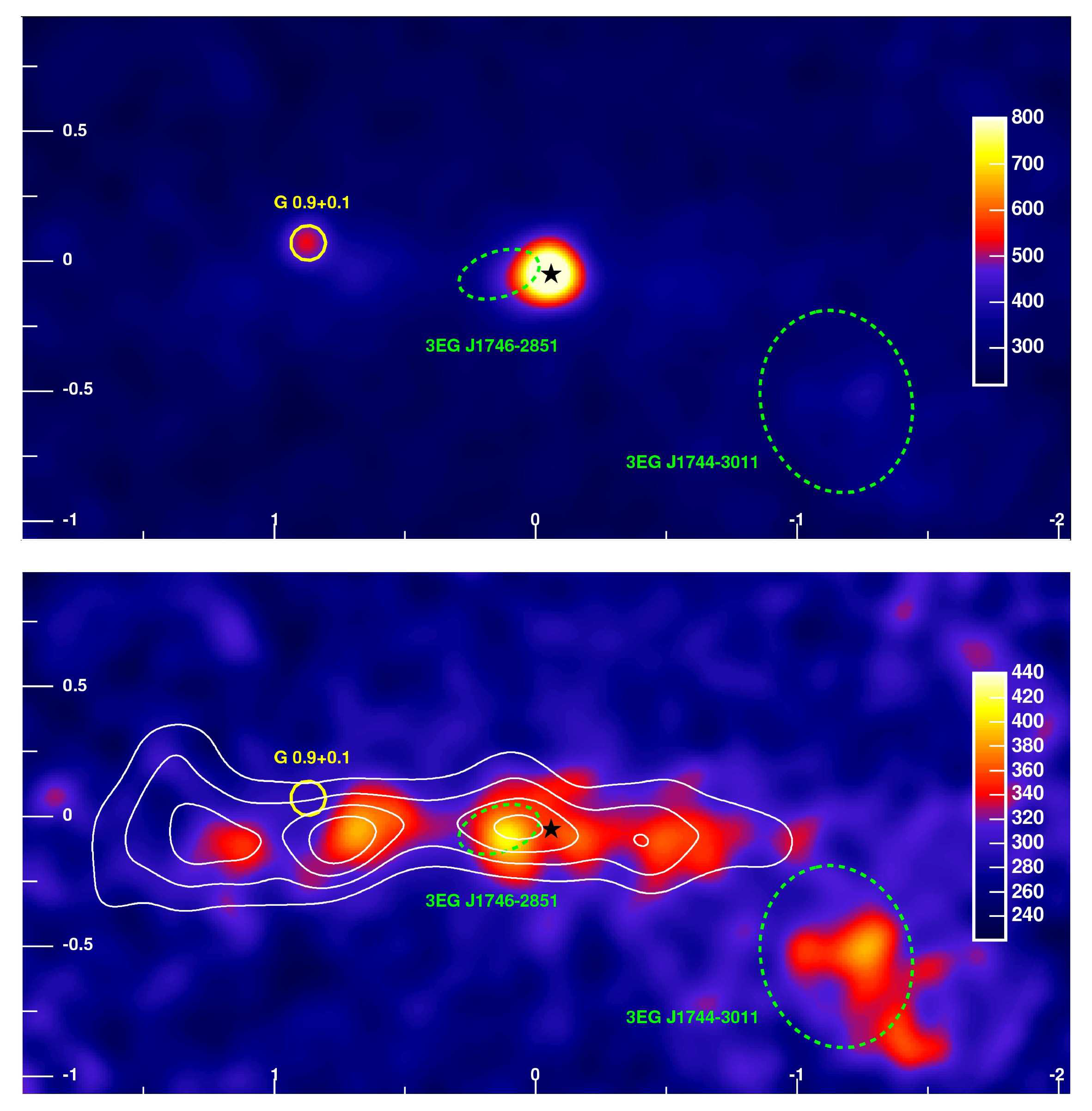}
  \caption{VHE gamma-ray images of the GC region. Top: gamma-ray count map, bottom: the same map after
subtraction of the two dominant point sources, showing an extended band of gamma-ray emission.
White contour lines indicate the density of molecular gas, traced by its CS emission. The position
and size of the composite SNR G0.9+0.1 is shown with a yellow circle. The position of Sgr A* is marked with a black star. Figure extracted from~\cite{2006Natur.439..695A}.}
  \label{viana:fig2}
\end{figure}
 Using 93h of data on the central source accumulated in the years 2004, 2005 and 2006, the energy spectrum of the gamma rays was measured with high precision, revealing an energy break or cutoff in the spectrum around 15 TeV (Fig. \ref{viana:fig1}, right). No signs of variability has been found~\citep{Aharonian:2009zk}. Different mechanisms have been suggested to explain the broadband spectrum of the GC. Firstly, the stochastic acceleration of electrons interacting with the turbulent magnetic field in the vicinity of Sgr A*, as discussed by \cite{Liu:2006bf}, has been advocated to explain the millimeter and sub-millimeter emission. This model would also reproduce the IR and X-ray flaring. In addition, it assumes that charged particles are accreted onto the black hole, and predicts the escape of protons from the accretion disk and their acceleration~\citep{Liu:2006bf}. These protons produce $\pi^0$ mesons by inelastic collisions with the interstellar medium in the central star cluster of the Galaxy. The cut-off energy found in the gamma-ray spectrum could reflect a cut-off E$_{cut,p}$ in the primary proton spectrum. In that case, one would expect a cut-off in the gamma-ray spectral shape at E$_{cut}$ $\simeq$ E$_{cut,p}$/30. The measured value of $\sim$15 TeV would correspond in this scenario to a cut-off energy in the primary proton spectrum between 100-400 TeV depending on the strength of the exponential cut-off. Energy-dependent diffusion models of protons to the outside of the central few parsecs of the Milky Way~\citep{Aharonian:2004jr} are alternative plausible mechanisms to explain the TeV emission observed with the H.E.S.S. instrument. They would lead to a spectral break as in the measured spectrum due to competition between injection and escape of protons outside
the vicinity of the GC.
\begin{figure}[t!]
 \centering
 \includegraphics[width=0.48\textwidth,clip]{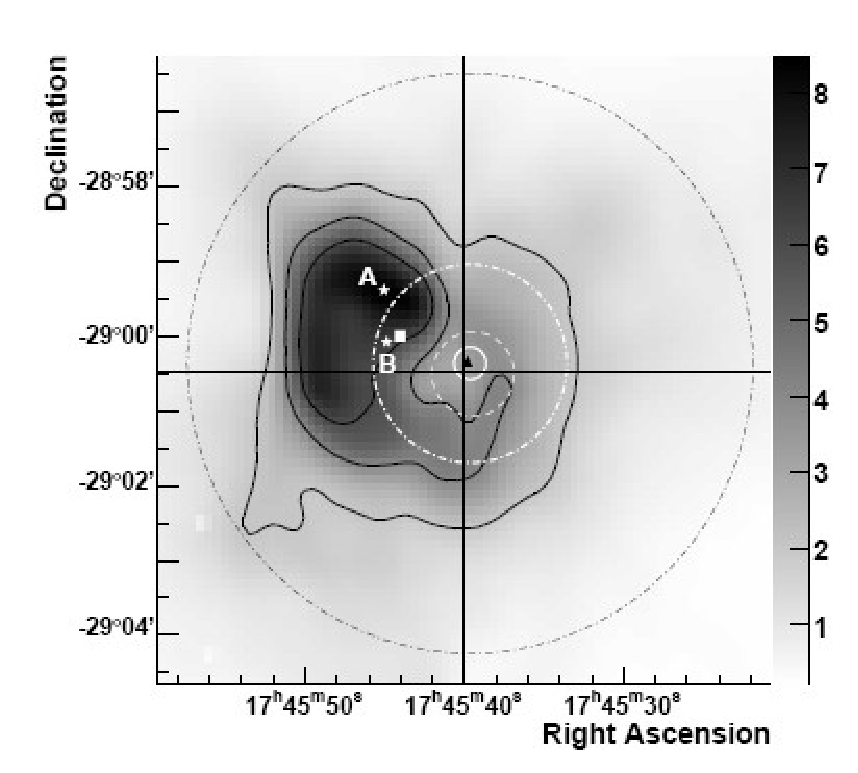}%
 \includegraphics[width=0.48\textwidth,clip]{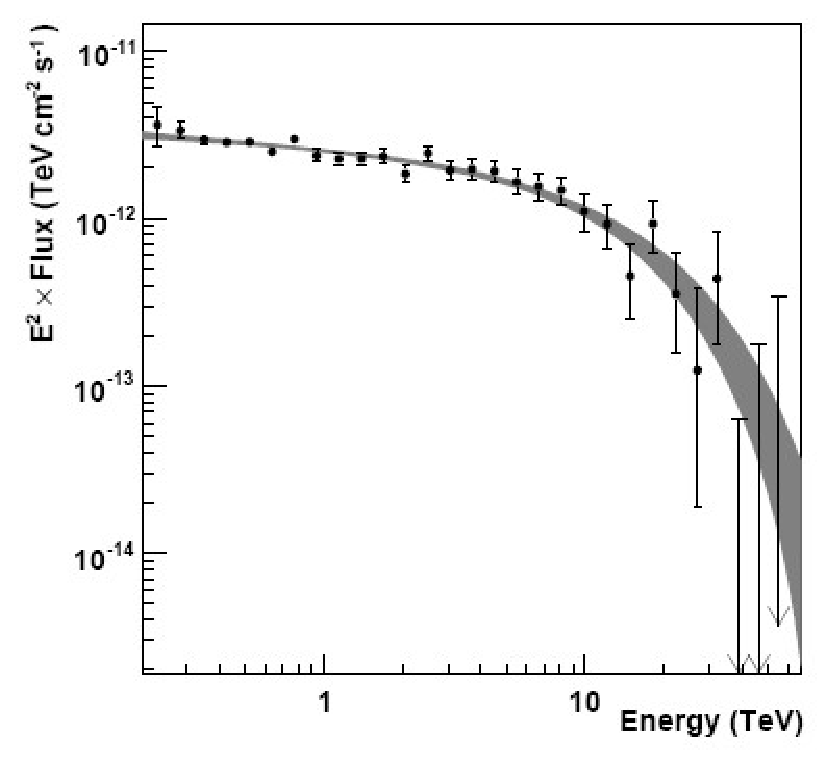}
  \caption{{\bf Left:} 90 cm VLA radio flux density map of the innermost 20 pc of the GC, showing emission from the supernova remnant Sgr A East. The crossing lines show the position of the Galactic Center Sgr A*. The 68$\%$ CL error contour for the position of the gamma ray source HESS J1745-290 is given by the small white circle. The white stars marked A and B denote the position of the radio maximum and the centroid of a radio emission from Sgr A East, respectively. Figure extracted from~\cite{2010MNRAS.402.1877A}. {\bf Right:} Energy spectrum of gamma rays from HESS J1745-290 as determined from the data sets obtained in the years 2004, 2005 and 2006. The shaded band shows the best fit to data for a power law with an exponential cutoff. Figure extracted from~\cite{Aharonian:2009zk}. }
  \label{viana:fig1}
\end{figure}

\section{The diffuse emission from the Galactic Center Ridge}

\begin{figure}[t!]
 \centering
 \includegraphics[width=0.45\textwidth,clip]{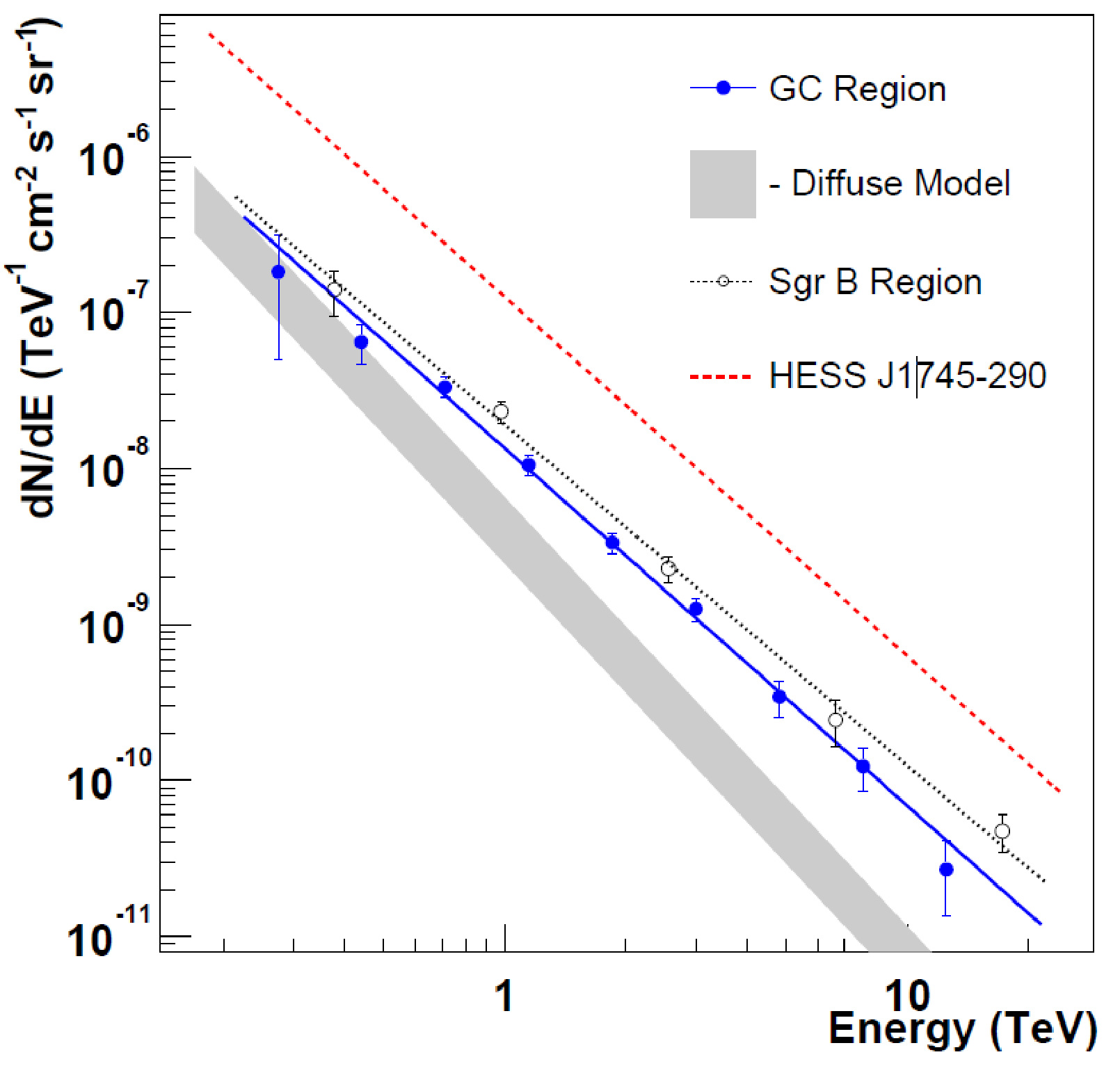}
  \caption{Gamma-ray flux per unit solid angle in the GC region (data points), in comparison with the
expected flux assuming a cosmic-ray spectrum as measured in the solar neighbourhood (shaded band). The strongest emission away from the bright central source HESS J1745-290 occurs close to the Sagittarius B complex of giant molecular clouds. The energy spectrum of this region is shown using open circles. The 2006 spectrum of the central source HESS J1745-290 is shown for
comparison (using an integration radius of 0.14$^{\circ}$). Figure extracted from~\cite{2006Natur.439..695A}.}
  \label{viana:fig3}
\end{figure}

In order to search for much fainter emission, an analysis of the GC region was made~\citep{2006Natur.439..695A} subtracting the best fit model for point-like emission at the position of HESS J1745-290 and the SNR G0.9+0.1. Two significant features are apparent after subtraction: extended emission spatially coincident with the unidentified EGRET source 3EGJ1744-3011 and emission extending along the Galactic plane for roughly 2$^{\circ}$. The latter emission is not only clearly extended in longitude l, but also significantly extended in latitude b (beyond the angular
resolution of H.E.S.S.) with a characteristic root mean square (rms) width of 0.2$^{\circ}$, as can be seen
in Fig.~\ref{viana:fig2}. The reconstructed gamma-ray spectrum for the region -0.8$^{\circ}$ < l < 0.8$^{\circ}$, |b| < 0.3$^{\circ}$ (with point-source emission subtracted) is well described by a power
law with photon index $\Lambda$ = 2.29 $\pm$0.07$_{\rm stat}$ $\pm$ 0.20$_{\rm sys}$ (Fig.~\ref{viana:fig3}).

Given the plausible assumption that the gamma-ray emission takes place near the center of the
Galaxy, at a distance of about 8.5 kpc, the observed rms extension in latitude of 0.2$^{\circ}$ corresponds
to a scale of $\approx$ 30 pc. This value is similar to that of interstellar material in giant molecular clouds
in this region, as traced by their CO emission and in particular by their CS emission~\citep{1999ApJS..120....1T}. At least for |l| < 1., a close match between the distribution of the VHE gamma-ray emission and the density of dense interstellar gas is found~\citep[see][for more details]{2006Natur.439..695A}. The close correlation between gamma-ray emission and available target material in the central 200 pc of our galaxy is a strong indication for an origin of this emission in the interactions of CRs. The hardness of the gamma-ray spectrum and the conditions in those molecular clouds indicate that the cosmic rays giving rise to the gamma-rays are likely to be protons and nuclei rather than electrons. Since in the case of a power-law energy distribution the spectral index of the gamma-rays closely traces the spectral index of the CRs themselves, the measured gamma-ray spectrum implies a CR spectrum near the GC with a spectral index close to 2.3, significantly harder than in the solar neighbourhood (where an index of 2.75 is measured). Given the probable proximity and young age of particle accelerators, propagation effects are likely to be less pronounced than in the Galaxy as a whole, providing a natural explanation for the harder spectrum which is closer to the intrinsic CR-source spectra. In addition, the key experimental facts of a harder than expected spectrum, and a higher than expected TeV flux, imply that there is an additional young component to the GC cosmic-ray population above the CR 'sea' which fills the Galaxy. This is the first time that such direct evidence for recently accelerated (hadronic) CRs in any part of our galaxy has been found.

The observation of a deficit in VHE emission at l = 1.3$^{\circ}$ relative to the available target material (see Fig.~\ref{viana:fig1}, bottom) suggests that CRs, which were recently accelerated in a source or sources in the GC region, have not yet diffused out beyond |l| = 1$^{\circ}$. Therefore the central CRs accelerators would only been active in the GC for the past 10,000 years. The fact that the diffuse emission exhibits a photon index which is the same - within errors - as that of HESS J1745-290 suggests that the underlying astrophysical emitter of HESS J1745-290 could be the source in question. Within the 1' error box of HESS J1745-290 are two compelling candidates for such a CR accelerator. The first is the SNR Sgr A East with its estimated age around 10 kyr. The second is the supermassive black hole Sgr A* which may have been more active in the past~\citep{Aharonian:2004jr}.

\section{Conclusions}
Observations with H.E.S.S. provide a
very sensitive view of this interesting region. With the recent data from the
H.E.S.S. instrument, a rich VHE gamma-ray morphology becomes evident, giving
strong indication for the existence of a cosmic ray accelerator within the central
10 pc of the Milky Way. Future observations with more sensitive instruments such as CTA will
significantly improve our knowledge about the GC region at VHE energies.


\bibliographystyle{aa}  
\bibliography{viana} 

\begin{thebibliography}{12}
\expandafter\ifx\csname natexlab\endcsname\relax\def\natexlab#1{#1}\fi

\bibitem[{{Acero} {et~al.}(2010)}]{2010MNRAS.402.1877A}
{Acero}, F. {et~al.} 2010, \mnras, 402, 1877

\bibitem[{Aharonian \& Neronov(2005)}]{Aharonian:2004jr}
Aharonian, F. \& Neronov, A. 2005, Astrophys.J., 619, 306

\bibitem[{Aharonian {et~al.}(2004)}]{Aharonian:2004wa}
Aharonian, F. {et~al.} 2004, Astron.Astrophys., 425, L13

\bibitem[{Aharonian {et~al.}(2005)}]{Aharonian:2005br}
Aharonian, F. {et~al.} 2005, Astron.Astrophys., 432, L25

\bibitem[{{Aharonian} {et~al.}(2006)}]{2006Natur.439..695A}
{Aharonian}, F. {et~al.} 2006, \nat, 439, 695

\bibitem[{Aharonian {et~al.}(2009)}]{Aharonian:2009zk}
Aharonian, F. {et~al.} 2009, Astron.Astrophys., 503, 817

\bibitem[{Albert {et~al.}(2006)}]{Albert:2005kh}
Albert, J. {et~al.} 2006, Astrophys.J., 638, L101

\bibitem[{Kosack {et~al.}(2004)}]{Kosack:2004ri}
Kosack, K. {et~al.} 2004, Astrophys.J., 608, L97

\bibitem[{Liu {et~al.}(2006)Liu, Melia, Petrosian, \& Fatuzzo}]{Liu:2006bf}
Liu, S.-M., Melia, F., Petrosian, V., \& Fatuzzo, M. 2006, Astrophys.J., 647,
  1099

\bibitem[{Schodel {et~al.}(2002)Schodel, Ott, Genzel, Hofmann, Lehnert,
  {et~al.}}]{Schodel:2002py}
Schodel, R., Ott, T., Genzel, R., {et~al.} 2002, Nature, 419, 694

\bibitem[{{Tsuboi} {et~al.}(1999){Tsuboi}, {Handa}, \&
  {Ukita}}]{1999ApJS..120....1T}
{Tsuboi}, M., {Handa}, T., \& {Ukita}, N. 1999, \apjs, 120, 1

\bibitem[{Tsuchiya {et~al.}(2004)}]{Tsuchiya:2004wv}
Tsuchiya, K. {et~al.} 2004, Astrophys.J., 606, L115

\end{thebibliography}

\end{document}